\documentclass[%
 reprint,
 amsmath,amssymb,
 aps,
]{revtex4-2}

\usepackage{graphicx}% Include figure files
\usepackage{dcolumn}% Align table columns on decimal point
\usepackage{bm}% bold math
\usepackage{soul}
\begin{document}

\preprint{APS/123-QED}

\title{Charge Transfer via Temporary Bonds in $C_{60} + C_{60}^+$ Molecular Collisions}
\author{J. Smucker}
 \affiliation{University of Connecticut, USA}
  \email{jonathan.smucker@uconn.edu}
\author{J. A. Montgomery Jr}
\affiliation{University of Connecticut, USA}
\author{M. Bredice}%
 \affiliation{University of Connecticut, USA}
 \author{M. G. Rozman}
\affiliation{University of Connecticut, USA}
\author{E. Yankson}%
 \affiliation{University of Connecticut, USA}
\author{R. C\^{o}t\'{e}}
\affiliation{%
University of Massachusetts Boston, USA}
\author{V. Kharchenko}
\affiliation{University of Connecticut, USA}
\affiliation{ITAMP, Center for Astrophysics$\,|\,$ Harvard \& Smithsonian, USA}

\date{\today}% It is always \today, today,

\begin{abstract}
We present a theoretical description of resonant charge transfer in collisions of nano-particles, specifically for $C_{60} + C_{60}^+$ collisions. We predict that transient bonds between colliding fullerenes can significantly extend the interaction time, allowing for a greater probability of charge transfer. In our model, the dumbbell-shaped $(C_{60}-C_{60})^+$ quasi-molecule, that is temporarily formed during the collision, is described as a dynamic system of 120 zero-range potentials. Using this model, we calculate the exchange interaction between colliding fullerenes and subsequently determine the corresponding charge transfer cross sections at different collision velocities. Our results have been verified with data obtained from quantum molecular dynamics simulations of the fullerene collisions. The presented theoretical model provides a description of the experimental data on the $C_{60} + C_{60}^+ $ resonant charge transfer collision through the inclusion of the temporary formation of dumbbell-shaped fullerene molecules at low collision velocities.
\end{abstract}

\maketitle

\section{Introduction}
Investigations of the charge transfer processes have a rich history due to its many applications in plasma physics, astrophysics, atmospheric science and chemistry \cite{AtomIonCollisionsReviewDelos,ColdAtomIonReviewMichal,CTReview}. In atom-ion systems, the physics of charge transfer collisions is well understood and the probabilities of underlying quantum processes can be determined with high accuracy {\it ab initio} calculations \cite{cote2016ultracold,Signatureswave,smith2014experiments}. A significantly higher level of complexity arises in the molecular charge transfer processes because the dynamics of the nuclear degrees of freedom may strongly influence the parameters of the exchange interaction between colliding particles. Since the advent of ultra-fast laser pulses, charge transfer collisions have also been used to probe fundamental quantum dynamics as these reactions are heavily dependent on the properties of the electronic wave function \cite{AttoCTReview}. Recently, charge transfer research has expanded to more complex systems such as large molecules, nano-size clusters, and condense matter materials \cite{los1990charge,ChargeTransferNanoTubesRao,ChargeTransferGraphene,CTDNA}. Fullerenes have become a popular molecule to study due to their high degree of symmetry \cite{campbell2000fullerene} and potential technological applications \cite{C60GrapheneCT,ultrafastNanoTubeC60CT,ChargeTransferNanoTubesRao}.

Total cross sections for $C_{60} + C_{60}^+ $ resonant charge transfer collisions were first measured by Rohmund and Campbell in 1997 \cite{SmallAngleC60Res}. Their experiment was arranged to detect charge transfer for small scattering angles. Expectation of the small angles was based on the analysis of numerous data on ion collisions with atoms and molecules. Glotov and Campbell later discovered $C_{60}$ resonant charge transfer at large scattering angles and used these results to update the originally published total cross section data \cite{LargeAngleC60Res}.

Several models \cite{SmallAngleC60Res,rappTheory,C60C70WWellTheory,Me} have been employed to explain the results of the small scattering angle measurements \cite{SmallAngleC60Res}. Charge transfer at large scattering angles implies that a noticeable portion of the charge transfer cross section is due to collisions where the $C_{60}$ molecules overlap. These kinds of reactions were not accounted for in any of the theories describing the small angle scattering cross sections. Such overlap leads to the temporary formation of the dumbbell $(C_{60}-C_{60})^+ $ quasi-molecule with additional chemical bonds between atoms from different $C_{60}$ molecules. These temporary bonds can act as "bridges"; creating efficient pathways for the charge transfer between fullerenes. They are also important for the transfer of the translational energy of fullerenes into their internal degrees of freedom. These processes bring complexity into developing a consistent theoretical model but they add the possibility of using measurements of charge transfer processes to probe collision dynamics of the internal structure of nanoparticles. Formation of stable fullerene dimers have been investigated in several articles \cite{C120C119Dynamics,C60dynmSemiempirical,C60dynmfusion}. We argue that the temporary formation of dumbbell fullerene molecules can create long lived compound states which increase the probability of charge transfer. The collision formation of the long lived dumbbell-shaped $(C_{60}-C_{60})^+ $ with different orientations have been confirmed in our Quantum Molecular Dynamics (QMD) simulations of slow fullerene collisions as well.

To illustrate the impact of the bridge formation on the charge transfer cross section in $C_{60} + C_{60}^+$ collisions, we employed a simplified model of the electronic structure of $C_{60}$ molecules. sixty zero-range electronic pseudo-potentials have been arranged in the geometry corresponding to the structure of $C_{60}$. The electronic energies, wave functions, and the exchange interaction between $C_{60}$ and $C_{60}^+$ have been calculated by combining these zero-range potential (ZRP) results with the Holstein-Herring method \cite{FluxPlaneTheory}. We computed the cross section of the charge transfer collisions $\sigma(v)$ as a function of the collision velocity $v$ and compared our results to experimental data. From the analysis of the experimental data, we predicted the life-time $\tau(v)$ of the $(C_{60}- C_{60})^+$ quasi-molecule created by the temporary chemical bonding between the $C_{60}$ and the $C_{60}^+$.

\section{methods}
To calculate charge transfer cross sections we start with the definition of a cross section for semi-classical atomic and molecular collisions with a central potential:
\begin{equation}\label{EqBasic}
    \sigma = \int^\infty_0 2 \pi b~ P(v,b) ~db,
\end{equation}
where $b$ is the impact parameter and $P(b,v)$ is the probability of charge transfer. For the resonance charge exchange process, $P(v,b)$ can be approximated as \cite{FluxPlaneTheory}:
\begin{equation}\label{Probability}
    P(v,b)= \sin^2 \Phi(v,b) = \sin^2 \left( {\int_{-\infty}^{\infty}\frac{\epsilon_g-\epsilon_u}{2}dt} \right),
\end{equation}
where $\epsilon_g$ and $\epsilon_u$ are the energies of the gerade and ungerade states of the $C_{60} + C_{60}^+$ system. Through the use of pseudo-potentials, we have developed a computationally inexpensive method for the analyses of fullerene collisions. In analogy to the charge transfer process in multi-electron atom-ion collisions, we assume that during $C_{60}+C^+_{60}$ collisions there is a single active electron which is bound by $N=120$ ZRPs with a time dependent distance between fullerenes $R(t)$. Each fullerene is modeled as 60 ZRPs. Individual ZRPs are centered on the carbon atoms of the fullerenes and each potential corresponds to a boundary condition imposed on the wave function of the active electron $\Psi({\bf r})$ at the location of the carbon atoms. This wave function is represented by a linear superposition of Green's functions \cite{ZRPText,ChibisovReview} centered on the carbon atoms: $ \Psi(r) =\sum_{n=1}^{N}c_i * G({\bf r}, {\bf R}_n,\epsilon)$, where ${\bf R}_n$ is the radius vector of the $n^{th}$ carbon atom and $\epsilon$ is the energy of the active electron. In our simplest model, Green's functions can be formally expressed using the normalized wave function $\psi_n ({\bf r})$ of an electron bound with energy $\epsilon$ by a single ZRP:

\begin{equation}
    \psi_n= \sqrt{ 2\pi \kappa} ~ G({\bf r}, {\bf R}_n ,\epsilon) = \sqrt{\frac{\kappa}{2 \pi}}~ \frac{e^{-\kappa|{\bf r}-{\bf R}_n|}}{|{\bf r}-{\bf R}_n|},\label{ZRPPsi}
\end{equation}
where $\kappa= \sqrt{ 2|\epsilon|}$ is related to the binding energy of the active electron in the field of a single ZRP \footnote[22]{The asymptotic wave function of $C_{60}$ electrons at large distances $r \gg R$ can be written using the asymptote of the Coulomb's Green functions and the spherical approximation for the $C_{60}$ cage symmetry $\psi (r,l=0) = A ~r^{\frac{1}{\kappa} -1} exp(-\kappa r)$ where $A$ is the normalization coefficient.}. Typically, ZRP systems are treated by solving the transcendental equation produced by the 120 boundary conditions \cite{ZRPText}. For a system of this size, however, accurate solutions to the transcendental equation are difficult to obtain. Instead, we can estimate the eigen energies by diagonalizing the Hamiltonian of the $C_{60}-C^+_{60}$ quasi-molecule, which is represented by the symmetric 120 $\times$ 120 matrix:
\begin{equation} \label{Mat}
H=\begin{bmatrix}
\epsilon_0 & \Delta_{1,2}(R_{1,2}) & \cdots & \Delta_{1,120}(R_{1,120})\\
\Delta_{1,2}(R_{1,2}) & \epsilon_0 & \cdots & \Delta_{2,120}(R_{2,120})\\
\vdots & \vdots & \ddots & \vdots\\ \Delta_{1,120}(R_{1,120}) & \Delta_{2,120}(R_{2,120}) & \cdots & \epsilon_0
\end{bmatrix},
\end{equation}
where $\Delta_{n,m}(R_{n,m})$ describes the exchange interaction between $\psi_n ({\bf r})$ and $\psi_m({\bf r})$ states localized on the $n^{th}$ and $m^{th}$ carbon atoms and $R_{n,m}= |{\bf R}_n -{\bf R}_m |$ is the distance between these atoms. As a crude estimate, this energy could be set to the negative of the ionization potential of a carbon atom. However, our approach used $\epsilon_0$ and the corresponding value of $\kappa$ as a fit parameter. The parameter $\kappa$ was adjusted so that the computed density distribution for the active electron in $C_{60}$ was similar to the electron density predicted by a density functional theory calculation. In this case, we can imagine the approximated wave function as an average of all the electrons in the $C_{60}$ molecule or as the wave function of a single positively charged hole.

\begin{figure}
    \includegraphics[scale=0.6]{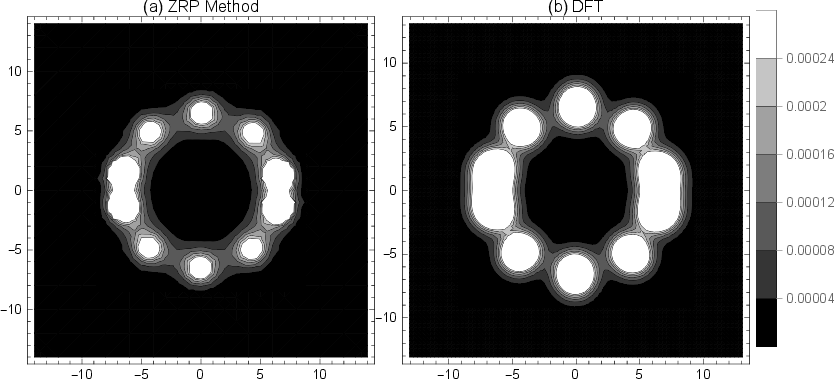}
    \caption{The electron wave function found using the zero range potential method (a) along side the electron density (normalized to one) calculated using density functional theory (b) for one plane. The zero range potential method used $k$ as a fit parameter and $\kappa=0.85$ Hartree$^{1/2}$. The spots of high probability correspond to carbon nuclei. The four spots closest to the x-axis are brightest since those carbon nuclei are the closest to the plane that is being plotted.}
    \label{Density}
\end{figure}

In order to calculate the exchange interaction $\Delta_{n,m}(R_{n,m})$, which describes the interaction between the $m$ and $n$ state, we use the Holstein–Herring method \cite{FluxPlaneTheory}. The exchange interaction between two degenerated states $\psi_n ({\bf r})$ and $\psi_m({\bf r})$ is expressed via the probability flux of the electron wave function through the plane that is perpendicular to ${\bf R}_{n.m}$ axis and half-way between the centers $m$ and $n$ \cite{FluxPlaneTheory}:
\begin{eqnarray}
\Delta_{n,m}(R_{n,m}) =2 \int^{\infty}_{-\infty}\int^{\infty}_{-\infty}~\psi_n(z=0)\nonumber\\
\times
\nabla \psi_m(z=0)\cdot \frac{\vec{R}_{n,m}}{||\vec{R}_{n,m}||}dy~dx,\label{Splitting}
\end{eqnarray}
choosing the plane that is halfway between the two potentials to be the x-y plane. When using the wave function shown in Eq.~(\ref{ZRPPsi}), Eq.~(\ref{Splitting}) yields the following result: $ \Delta_{n,m} = \kappa~{\rm exp}[-\kappa R_{n,m}/R_{n,m} ]$. Since this result is analytic, populating the matrix shown in Eq.(~\ref{Mat}), finding its eigenvalues and eigenvectors at different distances between the fullerenes can all be done at little computational cost. Fig.~\ref{Density} shows the electron density of a single $C_{60}$ molecule calculated using this method, alongside the electron density of a single $C_{60}$ molecule calculated using density functional theory and normalized to one electron. 

The computation of the charge transfer cross section in collisions between fullerenes requires an extended analysis of the collision trajectories ${\bf R}(t)$. In contrary with ion-atom collisions, where the straight line trajectory approximation provides an excellent description of the majority of experiments, trajectories of nanoparticles may be strongly influenced by the excitation of their internal structure and by the formation of long lived intermediate complexes. This change in trajectory becomes especially important for explaining the large scattering angles contribution to the total charge transfer cross section.

\section{Modeling Average Trajectory}
The collisional dynamics depend on the impact parameter and collision energy. Within the interval of experimental velocities reported in Glotov and Campbell \cite{LargeAngleC60Res} the fullerene collisions exhibit small scattering angles if the impact parameters are larger than the $C_{60}$ diameter $d= 13.33$ Bohr radii $a_0$ \cite{C60Radius}. This implies a relatively small value of the exchange phase $\Phi(v,b) < 1 $ and so a small probability of the charge transfer in Eq.~(\ref{Probability}), similar to what is usually observed in ion-atom collisions. If the impact parameter is less than diameter $d$, the average probability of charge transfer $\langle P(v,b)\rangle \simeq 1/2 $, due to the large value of the exchange interaction and exchange phase $\Phi(v,b)$. This reflects the efficient overlap of the internal structures of the colliding particles. The part of the charge transfer cross sections at $b < d$ is therefore estimated as $\sigma_c \simeq 0.5 \pi d^2$. Finally, when $b \sim d$ we expect the formation of temporary chemical bonds, or "bridges", between  fullerenes. These bridges and transfer of the collision energy into internal degrees of fullerenes extend the time of interaction between the particles at low collision velocities. Colliding fullerenes can form a long lived dumbbell $(C_{60}-C_{60})^+$ quasi-molecule. The life-time of this state and their structure depend on the collision velocity which affects the charge transfer probabilities between the fullerenes.
 
To visualize the mechanism of formation of temporary chemical bonds we performed QMD simulations of $C_{60} + C_{60}^+$ collisions at different velocities and impact parameters. Our simulations were performed using the atom centered density matrix propagation (ADMP) method \cite{schlegelMolDynm} as implemented in the Gaussian 16 electronic structure programs \cite{g16}. The electronic energy and its gradient are computed using the PM3 semi-empirical method \cite{stewart1,stewart2}.

\begin{figure}
    \includegraphics[scale=0.6]{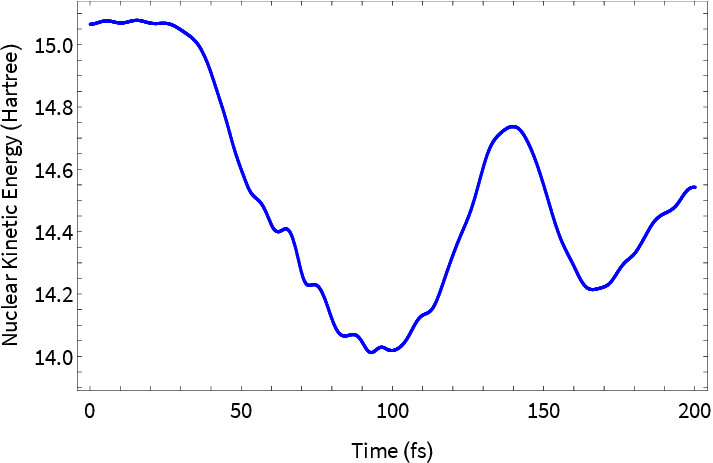}
    \caption{The total nuclear kinetic energy as a function of time for one of our simulations. This simulation had an impact parameter of $b=14.17 ~a_0$ and a collision energy of $410$~eV.}
    \label{KineticEnergy}
\end{figure}

The translation energy loss and collision time-delay have been investigated at different impact parameters and collision energies in the range from around $1$ eV up to several hundred eVs. Although the number of QMD simulations preformed is limited due to computational restraints. In Fig.~\ref{KineticEnergy} the kinetic energy of all 120 carbon atoms is shown as a function of time from one of our simulations. This simulation is at a collision energy of $409$ ~eV and an impact parameter of $b=14.17 ~a_0$, which is slightly large than the $C_{60}$ diameter $d$. In this relatively fast collision, the overlap of electronic states of different fullerenes is not enough to establish long lived bonds between the fullerenes and so only 3$\%$ of the translational kinetic energy is lost to the internal energies of the $C_{60}$ molecules. We found that the relative loss of translational energy is only significant for impact parameters below the diameter of $C_{60}$ and only slightly depends on collision velocities at higher collision energies. For collisions with an energy on the order of 1 eV, the interaction time between fullerenes is sufficiently long enough to establish "bridges" and the relative loss of the translational energy increases. The formation of "bridges" and the formation of the long lived dumbbell fullerene quasi-molecule have been observed at these lower collision velocities. The effective number of bridges and the time-delay associated with the formation of intermediate $(C_{60}-C_{60})^+$ state decrease with an increase of collision velocities and have been hardly seen at the velocities above $~ 2\times 10^4$ m/s.

Based on our QMD simulations and the results of other simulations \cite{C120C119Dynamics,C60dynmSemiempirical,C60dynmfusion}, the chemical bonds of $C_{60}$ molecules do not stretch significantly during collisions at the considered velocities. The molecules mostly experience temporary compression and temporary deformation during their close encounter. In order to include the compression of the molecules in our model, while we calculate the splitting between the gerade and ungerade as a function of the inter-molecular distance $R$, we forced the atoms to never be closer than $1$ {\AA}.

Key to our model is the prediction that temporary bonds between fullerenes significantly increase the probability of charge transfer. The time $\tau(v,b)$ of the interaction between two colliding $C_{60}$ fullerenes, the collisional time-delay, is the function of both the velocity $v$ and impact parameter $b$, and it can be determined  as:
\begin{equation}
\tau(E, b) = 2 \frac{d\eta_r(E,b)  }{dE} + 2 \frac{d\eta_c(E,b))  }{dE},
\end{equation}
where $\eta_r(E,b)$ and $\eta_c(E,b)$ are the Breit-Wigner resonance and background scattering phases respectively \cite{shimamura2011complete} and $E$ is the collision energy. During collisions involving fullerenes, the formation of dumbbell quasi-molecules occurs in various excited states with different configurations. Therefore, we expect several Breit-Wigner resonances arising at different resonance velocities $v_i$. The following simplified empirical formula can be used for the evaluation of the collisional time-delay and thus the life-time of the temporary chemical bonds:
\begin{equation}\label{Delay-Wigner}
\tau(v, b) = \sum_i {\rm a_{i}(b)} \frac{ \gamma_{i_c}}{(v^2-v_i^2)^2+\gamma_i^2} + \tau_c(v,b),
\end{equation}
where the first term is the sum of the Breit-Wigner resonances describing a formation of various long lived states of the quasi-molecule $(C_{60}-C_{60})^+$ with the total resonance width $\gamma_i(b)$. In multi-channel scattering, the partial width $\gamma_{i_c}(b)$ corresponds to the resonant charge transfer channel and the factor $a_i(b)$ takes into account the dependence of the relative strength of resonances on the impact parameter $b$. In the vicinity of resonance, the dependence of the background scattering time delay $\tau_c(v,b)$ on the velocity and impact parameter can be neglected. 

For the considered collision velocities, the trajectory of the particles are straight lines, $R(t) = \sqrt{b^2+v^2t^2}$, for most of the collision. Accurate trajectories need to take into account the time-delay $\tau(v,b)$ due to the temporary formation of quasi-molecules. The exponential decay of the quasi-molecules with the characteristic time $\tau(v,b)$, corresponds to the Breit-Wigner resonance formula. The average rate of change $v(t)$ of the interparticle distance can be represented as: $v(t)=v (1-e^{-|t|/\tau(v,b)})$, where $v$ is the collision velocity. This assumes that the same time is required for the collision formation as spontaneous decay of the quasi-molecule state due to the absolute value of the time being used. Therefore, the trajectories of $R(t)$ can be approximated by the following simple equation:
\begin{equation}
    R(t) = \sqrt{b^2+v^2t^2(1-e^{-|t|/ \tau(v,b)})^2},
\end{equation}
where $\tau (v,b)$ is the time constant that dictates how much the interaction time is extended by. This function approaches $v$ as $t$ approaches either negative infinity or positive infinity, meaning that at large distances $R$ the trajectory returns to the original straight line trajectory.

\section{Results}
\begin{figure}
    \includegraphics[scale=0.6]{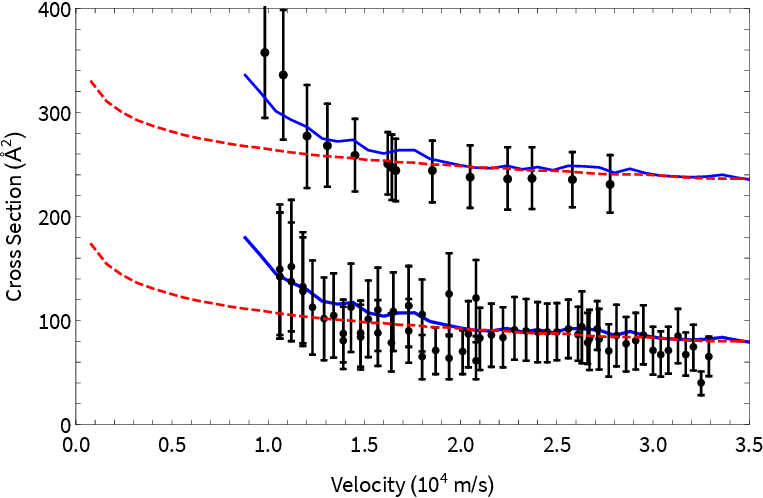}
    \caption{Data from Glotov and Campbell \cite{LargeAngleC60Res} (shown as black dots) along side our models. The upper part of the graph represents the total cross section and the lower part represents the small scattering angle cross section. The two solid curves are our model accounting for the extended interaction time due to temporary "bridge" formation. The two dashed curves represent our model without accounting for "bridge" formation.}
    \label{Final}
\end{figure}

The experimental data on the charge transfer cross sections\cite{LargeAngleC60Res} were measured for the interval of collision velocities $v > v_i$, where the time delay is not expected to be very large, and $\tau$ may be approximated using only a single resonance $v_i$ and no background time delay $\tau_{c}$. This single resonant velocity was fit to the data and found to be $9\times 10^3 m/s$. The width of the resonance $\gamma_i$ was also fit to the data and found to be $5 \times 10^2 m^2/s^2$. The factor $a_i(b)$ was taken to be $1/b^2$. $\gamma_{i_c}$ was fit as well and was $\gamma_{i_c}=300 a_0^2 m^4/s^3$. The effective size of the zero range potentials in our model was set to $\kappa=0.85$ Hartree$^{1/2}$. For collisions with $b < d$ we estimated charge transfer cross section as $\sigma_c \simeq {0.5}\pi d^2$. This core cross section should be not sensitive to the collision velocities  and it corresponds to the trajectories with large scattering angles and large recoil energy losses observed in the measurements of the fast neutral products \cite{LargeAngleC60Res}. The solid curves in Fig.~\ref{Final} shows the results of our model, which does account for the formation of quasi-molecules, for both the small scattering angle cross section (lower curve) and total charge transfer cross section (upper curve). Fig.~\ref{Final} also shows the experimental data for both the small scattering angle charge \cite{SmallAngleC60Res} transfer cross section and the total charge transfer \cite{LargeAngleC60Res} cross section along side different theoretical models. The lower dashed curve is the small scattering angle cross section computed using the electron wave function from the jellium model of the $C_{60}$ and the Holstein-Herring methods \cite{Me}. By assuming the probability of charge transfer is $1/2$ for all impact parameters below the diameter of $C_{60}$ $d$, this cross section was shifted up, shown as the upper dashed curve. This simplified model, which does extend the interaction time in order to account for the formation of quasi-molecules and does not fit well to the data at lower collision velocities. Results of our ZRP model calculations (the solid curves) are in very good agreement with the experimental data.

The available experimental data \cite{LargeAngleC60Res} can illustrate only a marginal increase of the cross sections due to suggested Breit-Wigner phenomena, i.e. the formation of the long-lived compound quasi-molecule in collisions of two fullerenes. We predict that the Breit-Wigner resonances at low collision velocities support long-lived bridges between fullerenes and this may lead to a significant increase of the charge transfer cross sections. The direct experimental measurements of the $C_{60} + C^+_{60}$ charge-transfer and scattering cross sections at energies of several eVs would be an important test of the bridge model and for the formation of the intermediate compound states in collisions of slow fullerenes.

\section{Conclusion}
The efficient computational model has been developed to explain experimental data on the cross section of $C_{60} + C_{60}^+$ resonant charge transfer collisions. In this model, the electronic state of the $C_{60}$ active electron  is described by an electron bound by $60$ identical zero range potentials evenly distributed across the $C_{60}$ molecule. The Hamiltonian of colliding fullerenes is reduced to the corresponding 120 $\times$ 120 matrix that describes the electronic state of the quasi-molecule formed in the $C_{60} + C_{60}^+$ collisions. The electronic states of the quasi-molecule are computed for the different time-dependent distance between colliding particles. The proposed model of the charge transfer process accounts for the formation of temporary chemical bonds between colliding fullerenes. These bonds significantly extend the interparticle exchange interaction. We predict that at lower collision energies the charge exchange between fullerenes is controlled by the Breit-Wigner resonances arising due to a temporary creation of the dumbbell quasi-molecule $(C_{60}-C_{60})^{*+}$. The formation of long lived quasi-molecule have been also observed in our QMD simulations. Results of our calculation of the resonant charge transfer cross section in $C_{60} +  C_{60}^+$ collisions are in good agreement with the experimental data.

\section*{Acknowledgments}
R. C was supported by the National Science Foundation (NSF) Grant No. PHY-2034284. V. K acknowledges support from the NSF through a grant for ITAMP at Center for Astrophysics $|$ Harvard \& Smithsonian.

\bibliography{references}% Produces the bibliography via BibTeX.

\end{document}